\documentclass[aip,apl,amsmath,amssymb,reprint,showpacs,showkeys,floatfix]{revtex4-1}

\usepackage{graphicx}
\usepackage{dcolumn}
\usepackage{bm}
\usepackage{siunitx}
\usepackage[belowskip=-5pt,aboveskip=0pt]{caption}
\begin{document}
\preprint{AIP/123-QED}

\title{Longitudinal spin-relaxation in nitrogen-vacancy centers in electron irradiated diamond}

\author{A. Jarmola}
\author{A. Berzins}%
\author{J. Smits}
\affiliation{Laser Centre, University of Latvia, 19 Rainis Boulevard, LV-1586 Riga, Latvia}
\author{K. Smits}
\affiliation{Institute of Solid State Physics, 8 Kengaraga Street, LV-1063 Riga, Latvia}
\author{J. Prikulis}
\affiliation{Institute of Chemical Physics, University of Latvia, 19 Rainis Blvd., LV-1586 Riga, Latvia}
\author{F. Gahbauer}
\author{R. Ferber}
\affiliation{Laser Centre, University of Latvia, 19 Rainis Boulevard, LV-1586 Riga, Latvia}
\email{ruvin.ferber@lu.lv}
\author{D. Erts}
\affiliation{Institute of Chemical Physics, University of Latvia, 19 Rainis Blvd., LV-1586 Riga, Latvia}
\author{M. Auzinsh}
\affiliation{Laser Centre, University of Latvia, 19 Rainis Boulevard, LV-1586 Riga, Latvia}
\author{D. Budker}
\affiliation{Helmholtz-Institut Mainz, Johannes Gutenberg Universit\"{a}t Mainz, 55128 Mainz, Germany}
\affiliation{Physics Department, University of California, Berkeley, Berkeley, CA 94720-7300, USA}

\date{\today}

\begin{abstract}
We present systematic measurements of longitudinal relaxation rates ($1/T_1$) of spin 
polarization in the ground state of the nitrogen-vacancy (NV$^-$) color center in 
synthetic diamond as a function of NV$^-$ concentration and magnetic field $B$. 
NV$^-$ centers were created by irradiating a Type 1b 
single-crystal diamond along the [100] axis with~\SI{200}{\kilo\electronvolt} electrons from a 
transmission electron microscope with varying doses to achieve spots of 
different NV$^-$ center concentrations. Values of ($1/T_1$)  
were measured for each spot as a function of $B$. 
\end{abstract}

\maketitle

\begin{figure}
  \centering
    \resizebox{\columnwidth}{!}{\includegraphics{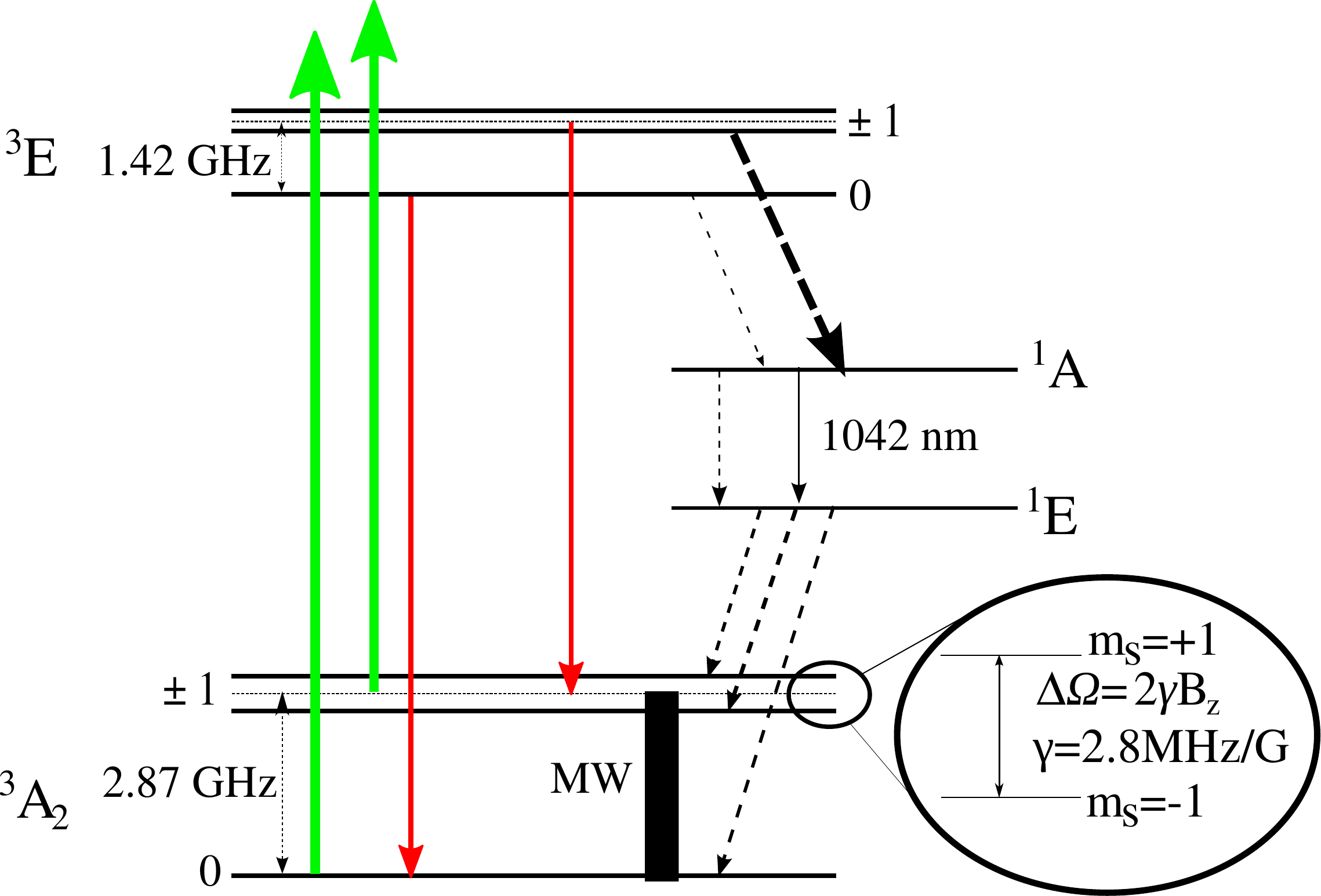}}
      \caption{\label{fig:levels}Energy level diagram of NV$^-$ center. Solid vertical lines 
      denote radiative transitions, whereas dashed lines denote transitions thought to be nonradiative.}
\end{figure}

Nitrogen-vacancy (NV$^-$) centers in diamond~\cite{Jelezko:2006} are 
useful for quantum information~\cite{GurudevDutt:2007}, 
magnetometry (see the review by Rondin \textit{et al.}~\cite{Rondin:2014}) 
and nanoscale sensing applications (see the review by 
Schirhagl \textit{et al.}~\cite{Schirhagl:2014}). NV$^-$ centers have been used to 
detect single electron spins~\cite{Grotz:2011,Mamin:2012,Shi:2015} and 
small ensembles~\cite{Mamin:2013,Staudacher:2013,Rugar:2014,Haberle:2015,DeVience:2015} 
and single nuclear spins~\cite{Sushkov:2015}, 
study magnetic resonance on a molecular scale, measure electric fields, strain and temperature, 
detect low concentrations of paramagnetic molecules and ions, and image 
magnetic field distributions of physical or biological systems.  
These applications are made possible by the unique properties of the NV$^-$ center 
level structure, shown in Fig.~\ref{fig:levels}, which allows manipulation of the 
ground-state spin state by optical fields and microwaves and measurement of the 
interactions of the ground-state spin with the local environment by monitoring the fluorescence 
intensity. Understanding spin relaxation processes is important in optimizing these techniques. 
Previous measurements of the dependence of longitudinal relaxation rate ($1/T_1$) of magnetic field 
have shown enhanced rates near $B=0$ G, $B=595$ G (Refs.~\cite{Armstrong:2010,Jarmola:2012,Anishchik:2015,Mrozek:2015}), 
and $B=514$ G (Refs.(~\cite{Jarmola:2012,Hall:2015}). The enhancements at zero field and 
$B=595$ G have been linked to interactions with NV centers whose orientation
makes their energies degenerate at these fields, 
while the enhancement at $B=514$~G is related to interactions with substitutional nitrogen 
(P1 centers). 
Previous work has also shown that the $1/T_1$ rate 
depends on NV concentration~\cite{Mrozek:2015}.
In this paper we describe systematic measurements of the longitudinal relaxation 
rates of ensembles of NV$^-$ centers created with different, controlled 
radiation doses on a single diamond crystal, achieved through irradiation with a transmission electron microscope (TEM). 
This method of preparing NV$^-$ centers is more convenient for many laboratories than irradiation in 
accelerators and also could facilitate the creation of microscopic structures on a diamond chip for special 
applications.

\begin{figure}
  \centering
    \resizebox{\columnwidth}{!}{\includegraphics{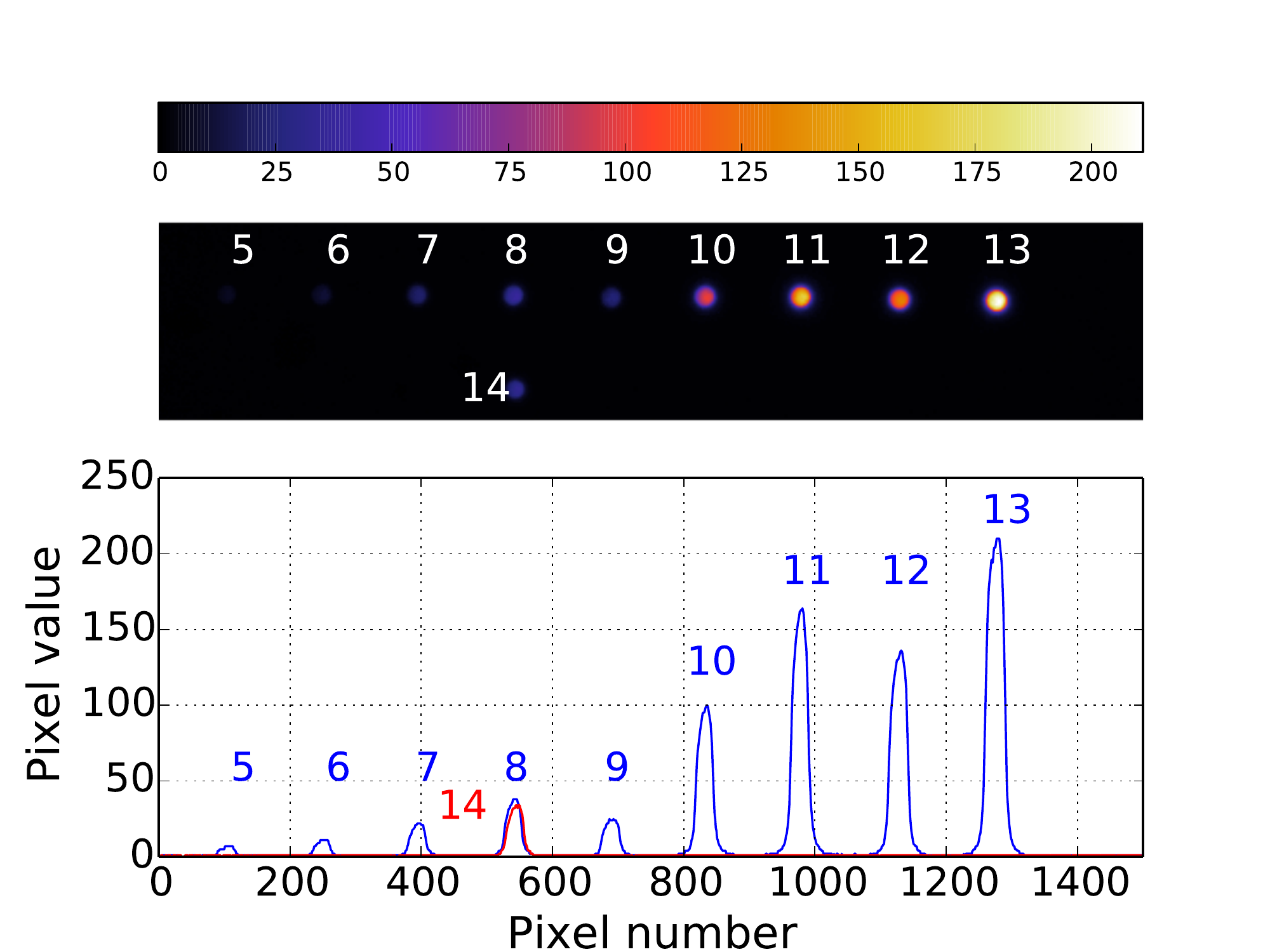}}
      \caption{\label{fig:luminosity}Fluorescence microscope image of the irradiated and annealed sample with a cross-section of fluorescence intensity. The colorbar indicates the intensity of the fluorescence.}
\end{figure}

The diamond sample was a [100] cut Type 1b single-crystal plate (Element 6), 
grown by the high-pressure, high-temperature technique, with an initial nitrogen concentration 
of 200 ppm and dimension of \SI{3}{\milli\metre} $\times$ \SI{3}{\milli\metre} $\times$ \SI{0.3}{\milli\metre}. 
It was irradiated using a 
transmission electron microscope~\cite{Kim:2012} (Tecnai G20 FEI with a Schottky field 
emitter electron source). The accelerating voltage was \SI{200}{\kilo\volt}. 
At these energies the electrons can be expected to penetrate about \SI{140}{\micro\metre} into the 
diamond\cite{estar}. However, it is unlikely that they will have sufficient energy to create vacancies below 
a depth of about ~\SI{20}{\micro\metre}\cite{Koike:1992}. 
Seventeen circular spots with a diameter of \SI{10}{\micro\metre} were irradiated. 
After irradiation, the sample was maintained at \SI{800}{\degreeCelsius} for three hours 
in the presence of nitrogen gas at less than atmospheric pressure. 
Then the sample was placed in a fluorescence microscope and illuminated with a mercury lamp through a filter cube
(Olympus U-MWG2), which reflected excitation light from~\SI{510}{\nano\metre} to \SI{550}{nm} to the sample, and passed the emission through 
a ~\SI{590}{\nano\metre} long-pass filter. The red fluorescence was photographed with a Peltier-cooled CCD camera (GT Vision GXCAM-5C). 
The pixel intensities of the fluorescence are shown in Fig.~\ref{fig:luminosity}; the 
irradiation parameters of the samples are given in Table~\ref{tab:irrad}.
Spots~5--13 were irradiated with an electron intensity of \SI{3530}{\per\nano\metre\squared\per\second}  
to reach the dose listed in the table. Spot~14 was obtained by irradiating 
with an electron intensity of \SI{2530}{\per\nano\metre\squared\per\second} to compare results for irradiation at 
different rates but same total intensity.

\begin{table}
\centering
\caption{Parameters of the irradiated spots numbered as per the labels in Fig.~\ref{fig:luminosity}: 
Integrated (red) fluorescence intensity upon irradiation with green light, electron dose and estimated 
NV$^-$ concentration.}
\label{tab:irrad}
\begin{tabular}{cccc}
\hline
\hline
    & \multicolumn{1}{c}{Integrated}		   	& \multicolumn{1}{c}{Electron}                      & Estimated\\
    & \multicolumn{1}{c}{Fluorescence}	   	& \multicolumn{1}{c}{Dose}                          & NV$^-$\\
Spot & \multicolumn{1}{c}{Intensity}		    &                                                   & Concentration\\
Nr.  & \multicolumn{1}{c}{(arb. units)}     & \multicolumn{1}{c}{(\si{\per\centi\metre\squared})} & (ppm) \\
\hline
5   & 1400						                      & $1.1 \times 10^{19}$                                & 0.2\\ 
6   & 2600 					   	                    & $2.1 \times 10^{19}$                                & 0.3\\ 
7   & 5700 					   	                    & $4.2 \times 10^{19}$                                & 0.7\\ 
8   & 10000					   	                    & $8.5 \times 10^{19}$                                & 1.2\\ 
9   & 6300 					   	                    & $1.7 \times 10^{20}$                                & 0.7\\ 
10  & 2900					   	                    & $3.4 \times 10^{20}$                                & 3.3\\ 
11  & 50000					   	                    & $6.8 \times 10^{20}$                                & 5.5\\ 
12  & 39000					   	                    & $1.3 \times 10^{21}$                                & 4.3\\ 
13  & 65000					   	                    & $2.5 \times 10^{21}$                                & 7.1\\ 
14  & 8300 					   	                    & $6.1 \times 10^{19}$                                & 3.9\\ 
\hline
\end{tabular}
\end{table}

\begin{figure}
\centering
\resizebox{0.9\columnwidth}{!}{\includegraphics{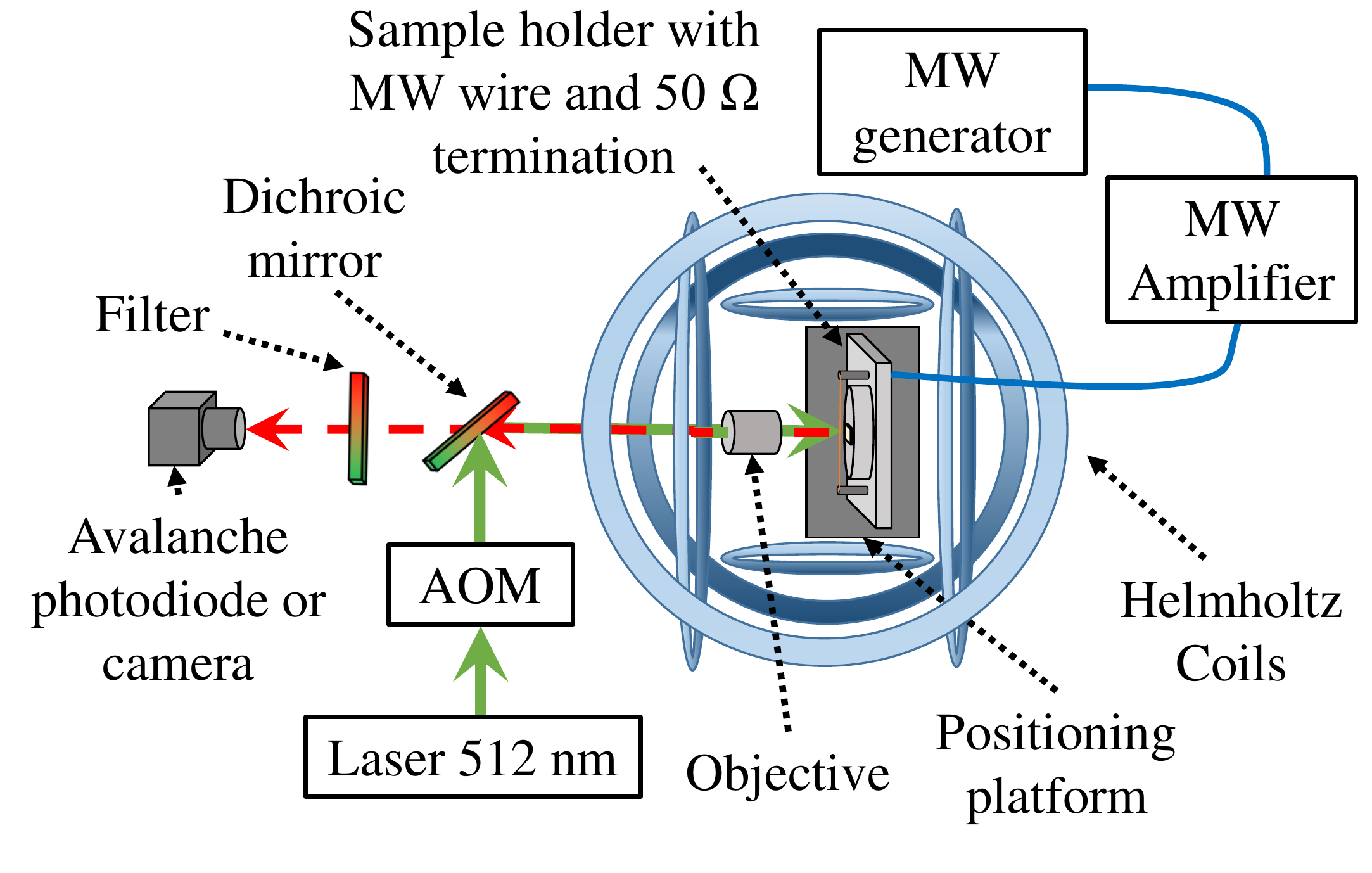}}
      \caption{\label{fig:exp}Schematic diagram of the experiment.}
\end{figure}

The experimental setup used to measure longitudinal relaxation times as a 
function of magnetic field is shown in Fig.~\ref{fig:exp}. 
One should note that our procedure assumes that the relaxation 
rates for the $m_S=0$ and $m_S=\pm1$ sublevels are identical.  
NV centers were excited with \SI{512}{\nano\metre} light from an
external cavity diode laser (Toptica DL100 pro). 
The laser power before the microscope lens was around \SI{7}{\milli\watt}. 
The microscope lens had a focal length of ~\SI{4.5}{\milli\metre} and numerical aperture of 0.55. 
The sample holder was mounted on a three-axis positioning stage (Thorlabs Max341) 
  and placed inside a three-axis Helmholtz coil system, allowing  
control over the magnetic field direction and magnitude. 
Microwaves (MW) from a frequency generator (Stanford Research Systems 
SG386) passed through a switch (Minicircuits ZASWA-2-50DR+) and were amplified with a 
\SI{16}{\watt} MW amplifier (Minicircuits ZHL-16W-43-S+). The MW were delivered to the sample by 
a \SI{0.071}{\milli\metre} diameter copper wire placed close to the diamond surface, and the 
wire was terminated into \SI{50}{ohm} after the sample.

The fluorescence from the sample was collected using the same microscope lens that 
focused the light onto the sample, and then passed through a dichroic mirror (Thorlabs DMLP567), 
which passed wavelengths longer than \SI{567}{\nano\metre}. 
After passing through an additional filter that further suppressed the green excitation light, 
the  fluorescence could be deflected 
to either a CMOS camera for visual adjustments of the sample, wire or position on sample, 
or to the avalanche photodiode (Thorlabs APD110A/M) for overall fluorescence measurements. 

\begin{figure}
\centering
\resizebox{0.7\columnwidth}{!}{\includegraphics{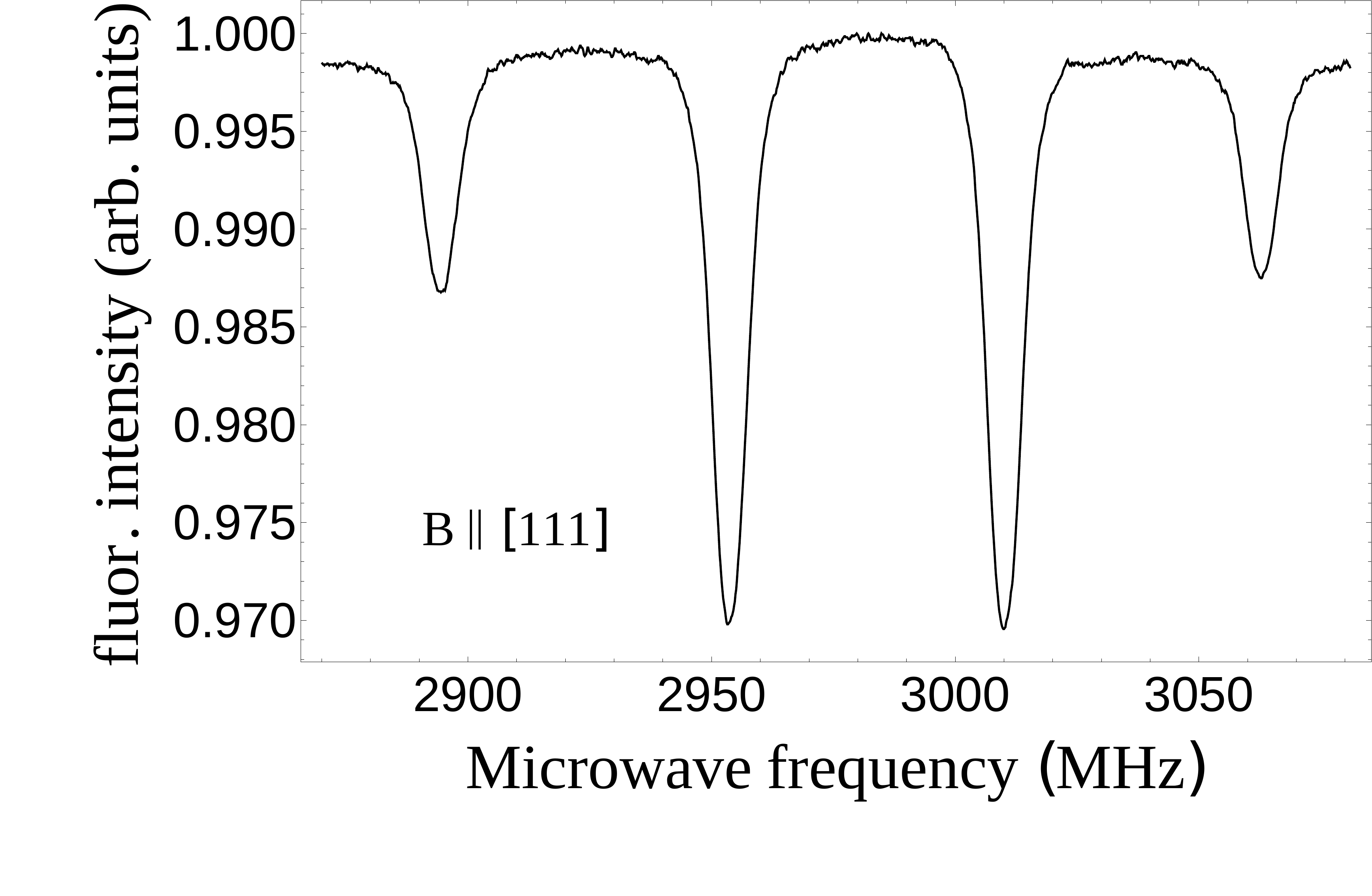}}
      \caption{\label{fig:odmr}Fluorescence vs. MW frequency at $B=30$ G.}
\end{figure}

The current in the Helmholtz coils was adjusted so that the magnetic field 
pointed in the [111] crystallographic direction. An optically detected 
magnetic resonance (ODMR) signal was obtained by scanning the microwave 
frequency under continuous laser irradiation and measuring the fluorescence (see Fig.~\ref{fig:odmr}). 
In this configuration two ODMR peaks appear on either side of the 
microwave frequency of \SI{2.87}{\giga\hertz}, which is the frequency of the NV resonance in the 
absence of magnetic field. The two inner peaks are more intense and 
correspond to the three possible alignments of the NV axis that make 
the same angle with the magnetic field. The two outer peaks correspond to the 
alignment of the NV axis that is parallel to the magnetic field. 
Then, the microwave generator was set to the frequency corresponding 
to one of the outer ODMR peaks (both were measured).  

The laser light hitting the sample could be turned on and off with an 
accousto-optic modulator (AA OptoElectronic MT200-A0.5-VIS). 
An pulse generator (PulseBlaster ESR-PRO-500)
was used to control the accousto-optic modulator and MW switch. To measure the longitudinal relaxation 
time, a decay curve was generated as follows. First, the NV$^-$ centers were pumped into  
$m_S=0$ level of the ground state with a green laser pulse that lasted~\SI{50}{\micro\second}. Then 
the laser light was blocked for a variable time $\tau$, after which the sample was again illuminated. 
The fluorescence as a function of time was recorded on an oscilloscope (Yokogawa DL6154), averaged 
1024 times and saved to disk. 
The procedure was repeated, but a 
microwave $\pi$ pulse was applied after the laser pump pulse. This sequence was repeated five times.  
The fluorescence 
obtained after time $\tau$ with the $\pi$ pulse was subtracted from the fluorescence obtained 
after time $\tau$ without the $\pi$ pulse to eliminate contributions to the signal from other 
NV$^-$ alignments and other sources of common-mode noise~\cite{Jarmola:2012}. The fluorescence immediately after 
time $\tau$ was normalized to the fluorescence after the spins have been pumped into the $m_S=0$ ground-state 
sublevel, and this quantity $I$ was plotted as a function of $\tau$.
The plot was fit with a stretched exponential of the form $\exp{-(\frac{\tau}{T_{1}})^{\beta}}$, 
where $T_{1}$ is the longitudinal relaxation time, and $\beta$ is a parameter between zero and one that 
describes the distribution of relaxation times in a large ensemble of NV$^-$ centers with similar but not identical 
relaxation times. A value of $\beta=1$ indicates a $\delta$-function distribution of relaxation times~\cite{Johnston:2006}.

\begin{figure}
  \centering
      \resizebox{0.9\columnwidth}{!}{\includegraphics{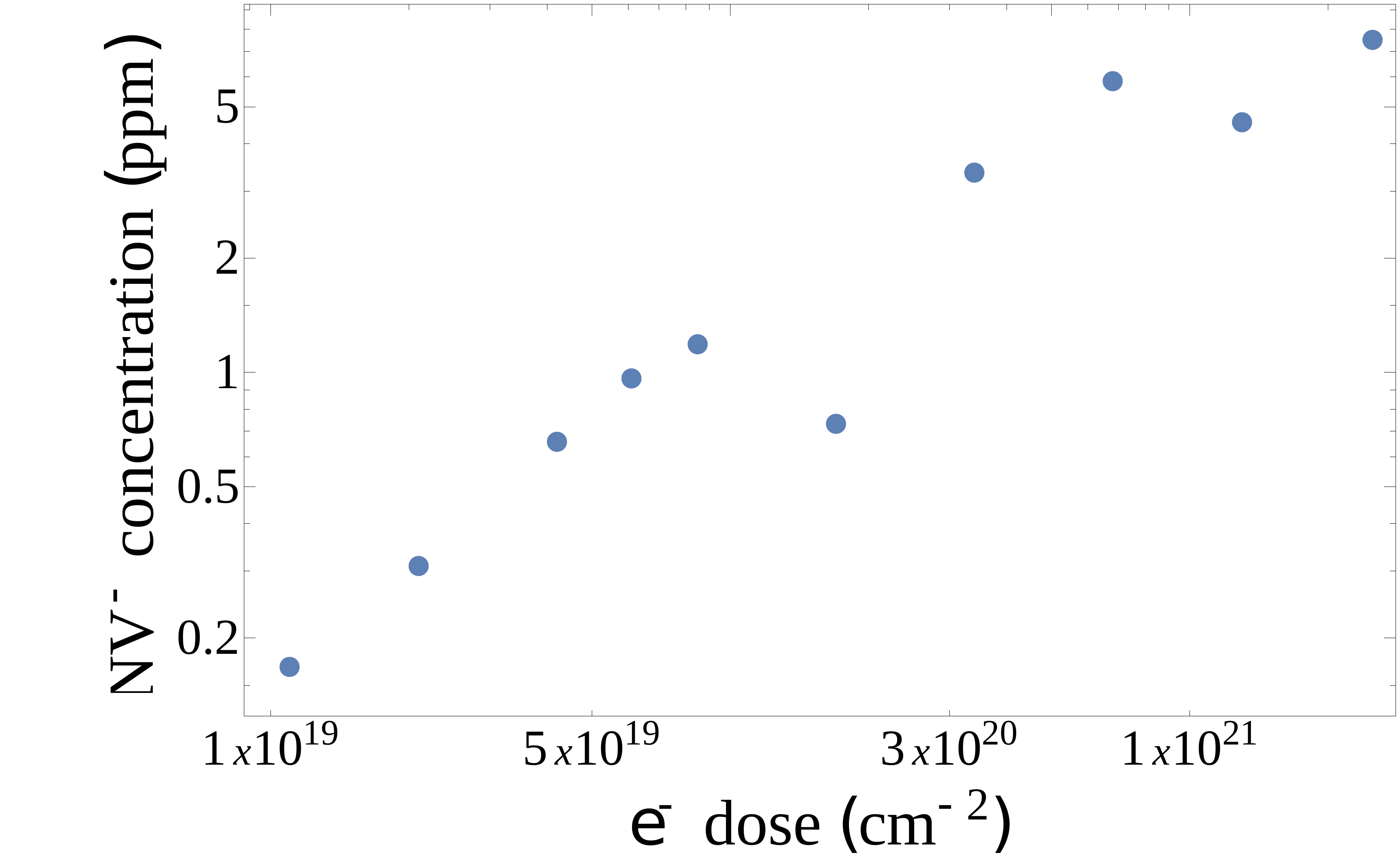}}
      \caption{\label{fig:concentration} Estimated NV$^-$ concentration vs electron dose. Error bars are not 
      given as the main error is systematic and could be a factor of five.}
\end{figure}

In order to compare the NV concentrations of the different spots, we measured 
their fluorescence with a fluorescence microscope (see Fig.~\ref{fig:luminosity}). 
The relationship between the fluorescence intensity and NV$^-$ concentration was determined using a diamond 
sample with a known, uniform NV$^-$ concentration of 10 ppm in the same setup. Assuming that 
the fluorescence intensity is proportional to the NV$^-$ concentration, the relative 
concentration of our spots should be relatively well known (see Fig.~\ref{fig:concentration}). 
However, the overall normalization should be 
considered to be only an order-of-magnitude estimate. Our estimated NV$^-$ concentrations are systematically 
lower than concentrations quoted for similar electron doses~\cite{Acosta:2010,Jarmola:2012,Mrozek:2015}; however, 
these experiments used higher electron energies. 

\begin{figure}
    \centering
      \resizebox{0.9\columnwidth}{!}{\includegraphics{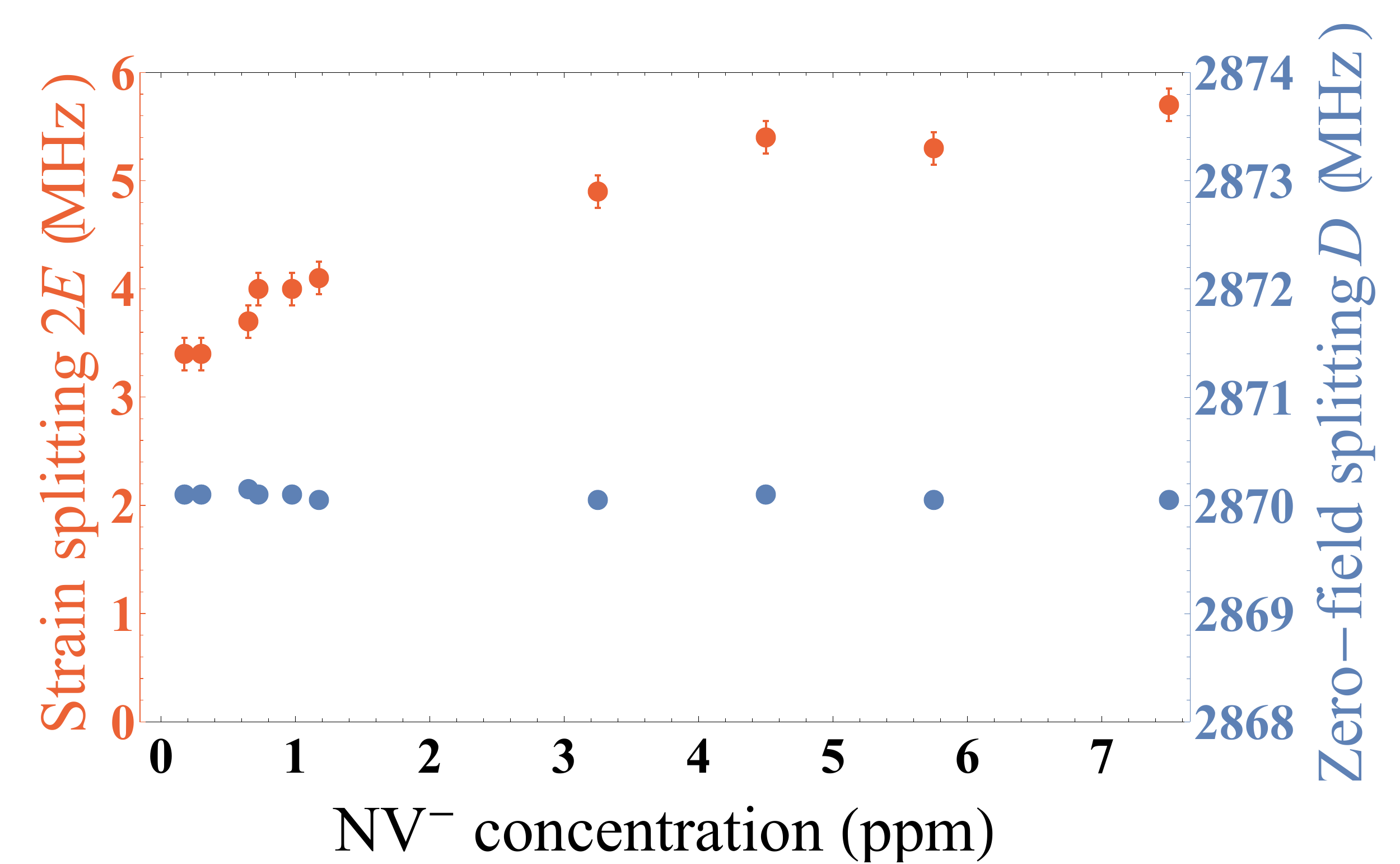}}
      \caption{\label{fig:strain} Strain splitting ($2E$) and zero-field splitting ($D$) as a function of 
      NV$^-$ concentration.}
\end{figure}

The Hamiltonian $H_S$ of the $^3A_2$ state (see Fig.~\ref{fig:levels}) of the electronic spin of 
the NV$^-$ center is given by
$H_S= DS_z^2 + E(S_x^2-S_y^2)+g_S \mu_B \vec{B}\cdot\vec{S}$,
where $DS_z^2$ corresponds to the zero-field splitting,  $E(S_x^2-S_y^2)$ corresponds to electric fields, which 
can arise in the lattice due to strain, and $g_S\mu_B\vec{B}\cdot\vec{S}$ corresponds to the Zeeman splitting. 
The strain parameter $2E$ can be observed as a small splitting in the zero-field ODMR signal. We have estimated this 
parameter for each spot by fitting parabolas to the peaks, 
and it is plotted in Fig.~\ref{fig:strain} as a function of estimated concentration. 
The radiation damage causes distortions in the crystal structure, as shown by the fact that the strain 
splitting increases with concentration, though more slowly above $~3$~ppm. The distortion could also change 
the distance between the N and V defects, which would alter the zero-field splitting $D$. 
Whereas an earlier study reported an increase of the zero-field 
splitting $D$ on the order of 20 MHz at high radiation doses~\cite{Kim:2012}, our measurements show no change in $D$.

\begin{figure}
  \centering
    \resizebox{\columnwidth}{!}{\includegraphics{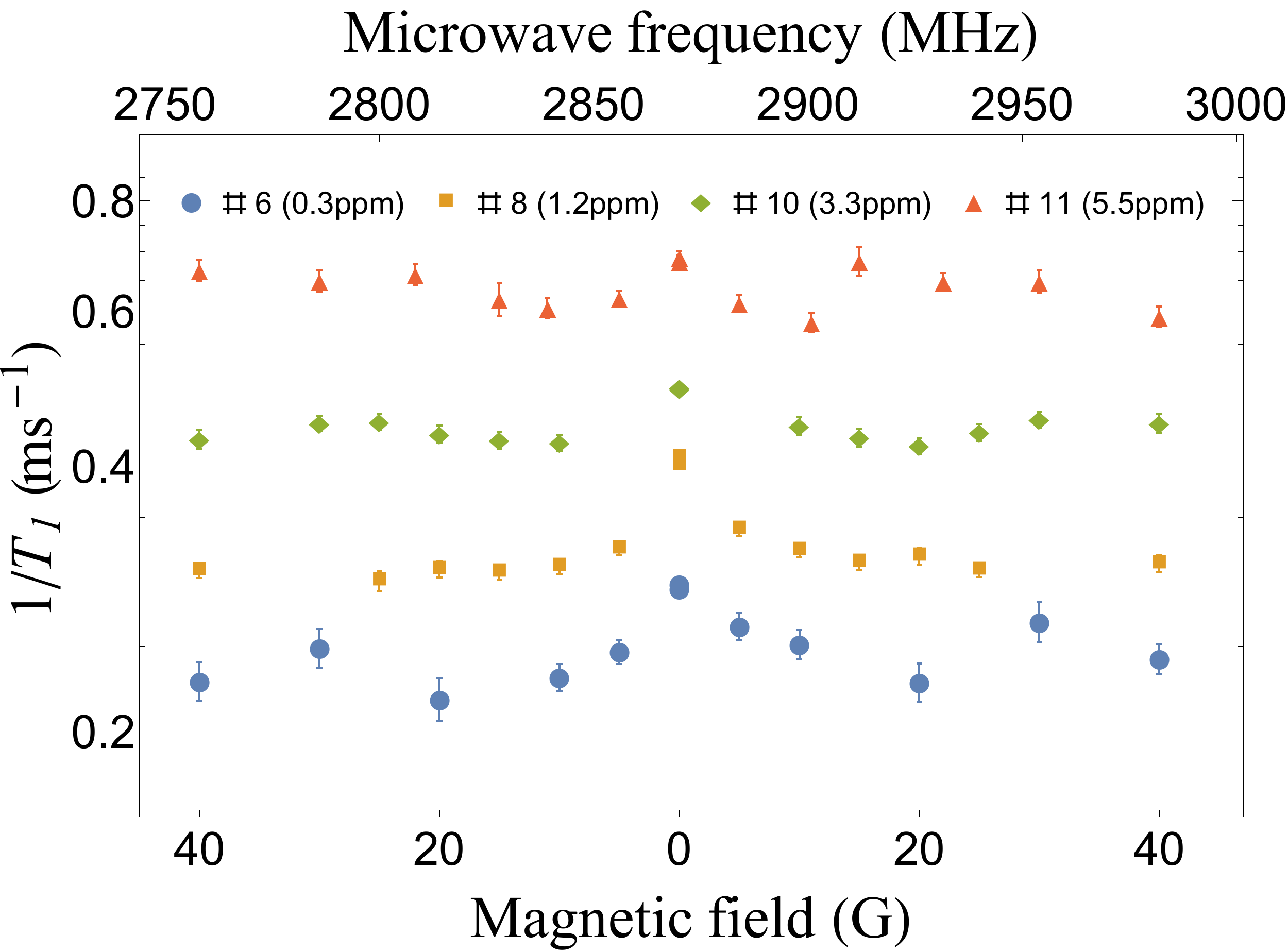}}
      \caption{\label{fig:rates}Longitudinal relaxation rates as a function 
      of magnetic field for different spots.}
\end{figure}

Figure~\ref{fig:rates} shows the measured longitudinal relaxation rates, obtained by fitting the decay curves from our 
experiments, as a function of magnetic field for various spots. 
The rates are comparable to previously measured rates in bulk samples~\cite{Jarmola:2012,Mrozek:2015}.  
The increase in $1/T_1$ at zero magnetic field is caused by the fact that all ODMR components are degenerate there\cite{Mrozek:2015}. 
The effect is analogous to level-crossing resonances. 
It is also possible to see a hint of the relaxation rates dropping and increasing again as the magnetic field value increases 
from zero, which is related to partial overlapping of the ODMR components at lower magnetic field values\cite{Mrozek:2015}.
More measurements near zero field would be desirable, but we could not resolve the $\pm1$ ODMR peaks at lower fields. 

\begin{figure}
  \centering
    \resizebox{0.9\columnwidth}{!}{\includegraphics{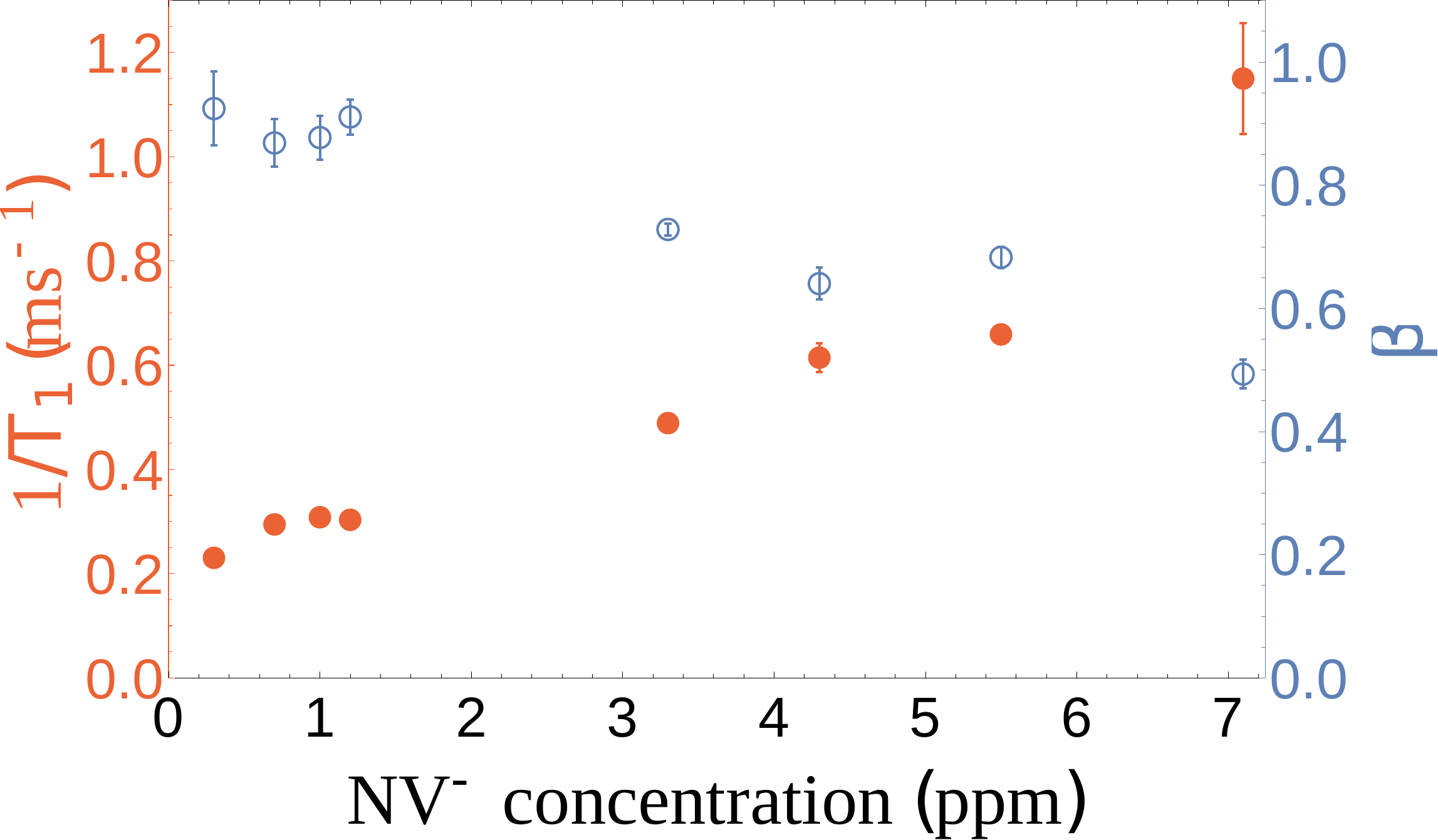}}
      \caption{\label{fig:beta_t1} $1/T_1$ (filled circles) and the 
      parameter $\beta$ (open circles) as a function NV$^-$ concentration at $B=30$~G.}
\end{figure}

Figure~\ref{fig:beta_t1} shows that in our fits the value of $\beta$ approaches unity as the NV$^-$ concentration decreases. 
This result is consistent with the measurements in Fig. 2(a) of~\cite{Jarmola:2012}, which showed that at low temperatures, 
the NV$^-$ density, i.e., interaction with nearby spins, dominated the $1/T_1$ rate, 
wherease at higher temperatures, relaxation induced by phonons dominated. 
Similarly, the nearly linear relationship between longitudinal relaxation rate and NV concentration 
is consistent with dipole-dipole interactions driving relaxation at these densities, 
since the dipole field drops off as $r^{-3}$ whereas the average distance to the nearest NV center changes as $\rho^{-1/3}$.

In conclusion, we have made a systematic measurement of longitudinal relaxation times as a function of magnetic field and 
and NV$^-$ concentration for a type Ib diamond.  
The results suggest that at high concentrations the NV$^-$ centers in the spots do not have a narrowly defined relaxation rate, 
but rather some distribution of relaxation rates. 
However, as the concentration decreases, the relaxation rate becomes 
better defined, as indicated by the stretched-exponential fits of the polarization-decay curves 
approaching unity. 
Another interesting feature is the zero-field resonance in relaxation rate when plotted against magnetic 
field (Fig.~\ref{fig:rates})~\cite{Jarmola:2012,Mrozek:2015}.
Finally, the longitudinal relaxation rate was measured for values of NV$^-$ concentration spanning an order of magnitude, 
and the value of the relaxation rate increased by almost a factor of five over this range.  
These measurements could help to guide the preparation of microscale NV$^-$ sensors on diamond using a TEM.

This work was supported by ESF Project Nr. 2013/0028/1DP/1.1.1.2.0/13/APIA/VIAA/054. 
We thank Raimonds Poplausks and Kaspars Vai\v{c}ekonis for help with the experiments. 
D.B. acknowledges support by DFG through the DIP program (FO 703/2-1) and by the
AFOSR/DARPA QuASAR program. 

\bibliography{nvbib}

\end{document}